\documentclass[journal=jacsat,manuscript=article]{achemso}

\usepackage[version=3]{mhchem}
\usepackage{graphicx}
\usepackage{dcolumn}
\usepackage{physics}
\usepackage{gensymb}
\usepackage{bm}
\usepackage{xcolor}
\usepackage{footmisc}
\usepackage{longtable}
\usepackage{xr-hyper}
\usepackage{hyperref}
\usepackage{chemformula} 
\usepackage{textgreek}

\author{A.~Gardill}
\author{I.~Kemeny}
\author{M.~C.~Cambria}
\author{Y.~Li}
\affiliation{Department of Physics, University of Wisconsin, Madison, Wisconsin 53706, USA}

\author{H.~T.~Dinani}
\author{A.~Norambuena}
\affiliation{Centro de Investigaci\'on DAiTA Lab, Facultad de Estudios Interdisciplinarios, Universidad Mayor, Santiago, Chile}

\author{J.~R.~Maze}
\affiliation{Instituto de F\'isica, Pontificia Universidad Cat\'olica de Chile, Casilla 306, Santiago, Chile}
\alsoaffiliation{Centro de Investigaci\'on en Nanotecnolog\'ia y Materiales Avanzados, Pontificia Universidad Cat\'olica de Chile, Santiago, Chile}

\author{V. Lordi}
\affiliation{Lawrence Livermore National Laboratory, Livermore, CA, 94551, USA}

\usepackage{etoolbox}
\usepackage{blindtext}

\makeatletter
\patchcmd{\acs@contact@details}{E}{*\,E}{}{}
\makeatother
\author{S.~Kolkowitz}
\email{kolkowitz@wisc.edu}
 
\affiliation{Department of Physics, University of Wisconsin, Madison, Wisconsin 53706, USA}

\title{Probing charge dynamics in diamond with an individual color center}

\begin{document}
\begin{abstract}
Control over the charge states of color centers in solids is necessary in order to fully utilize them in quantum technologies. However, the microscopic charge dynamics of deep defects in wide-bandgap semiconductors are complex, and much remains unknown. Here, we utilize single shot charge state readout of an individual nitrogen-vacancy (NV) center to probe charge dynamics of the surrounding defects in diamond.
We show that the NV center charge state can be converted through the capture of holes produced by optical illumination of defects many microns away. With this method, we study the optical charge conversion of silicon-vacancy (SiV) centers and provide evidence that the dark state of the SiV center under optical illumination is SiV$^{2-}$. These measurements illustrate that charge carrier generation, transport, and capture are important considerations in the design and implementation of quantum devices with color centers, and provide a novel way to probe and control charge dynamics in diamond. 
\end{abstract}

Defects in diamond such as nitrogen-vacancy (NV) \cite{Doherty2013} and silicon-vacancy (SiV) \cite{Muller2014} centers have emerged as promising quantum science platforms \cite{Rondin2014, Dolde2011,Togan2010, Dolde2014, Evans2018, Bradley2019, Bhaskar2020}. However, control over the charge states of color centers is critical to realizing many applications. As a result, techniques for preparing color center charge states are actively being explored \cite{Waldherr2011, Grotz2012, Fu2010, Hauf2011, Rose2018, Aslam2013, Irber2021}, but much remains unclear, and to date experiments have primarily focused on the charge state of individual centers. A deeper understanding of the dynamics 
of charge carriers in diamond is required in order to move beyond isolated color centers to arrays \cite{Dolde2014, Bradley2019, Oberg2019}.

\begin{figure}
\includegraphics[width=0.96\textwidth]{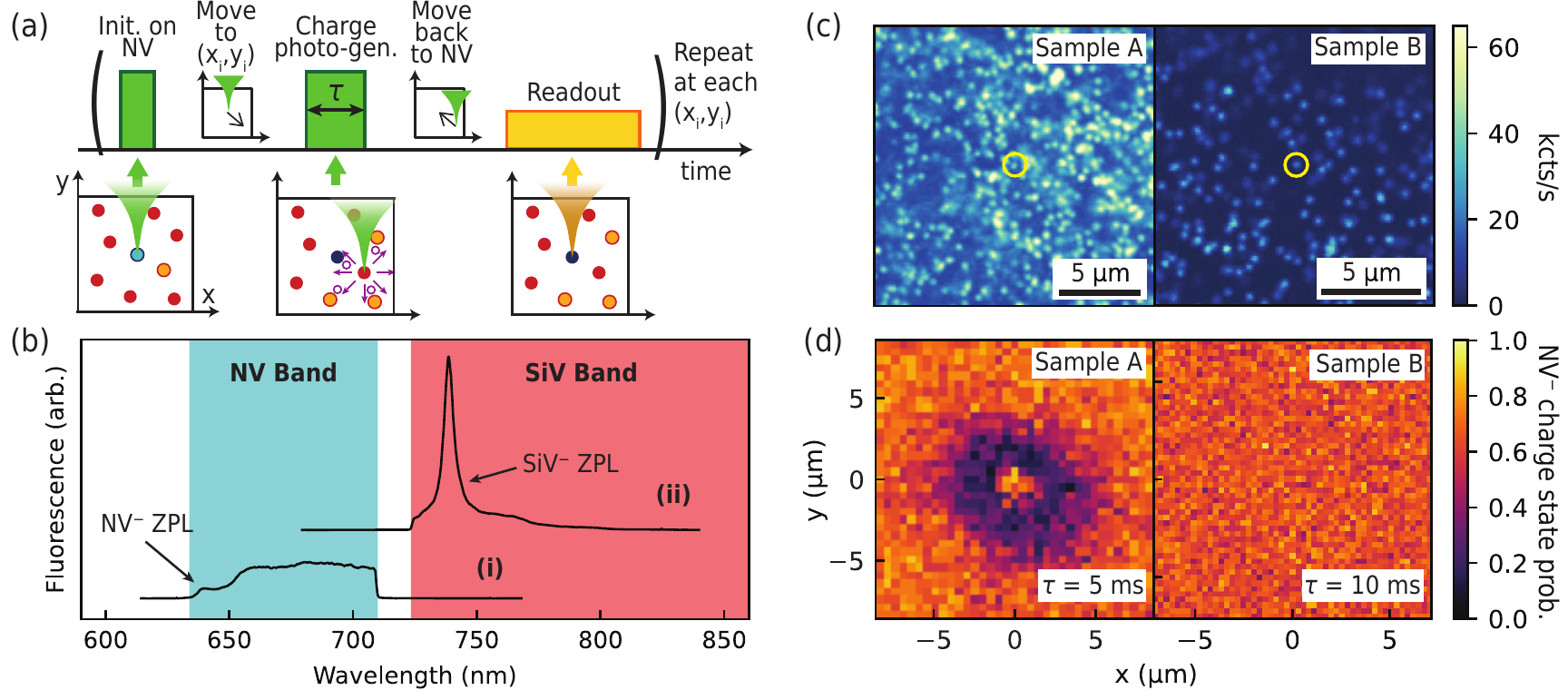}
\caption{\label{fig:SPaCE_measurement} Spatial Photogeneration and Capture of chargE (SPaCE) measurement 
(a) Pulse sequence of SPaCE measurement scheme, showing optical initialization (515~nm, 2~mW, 1~\textmu s), charge photogeneration (CPG)  (515~nm, 2~mW, $\tau$), and readout pulses (589~nm, 3~\textmu W, 100~ms), and movement of microscope focus between each pulse. Cartoon panels below show various defects: NV$^-$ (light blue circle), NV$^0$ (dark blue circle), SiV$^-$ (yellow circle), and SiV$^{2-}$ (red circle), undergoing the photogeneration of charge carriers (purple circle), and the transport and capture of these charges over the course of the sequence.  
(b) Optical spectra showing (i) NV$^-$ zero phonon line (ZPL) and phonon side band (515~nm excitation, 4~mW, 50~s integration), and (ii) SiV$^-$ ZPL (589~nm excitation, 0.34~mW, 180~s). Spectrum (i) was taken in a different sample with a higher density of NV centers for clarity of presentation. Blue and red regions indicate the range of the filters used to isolate the NV spectral band (630~nm to 715~nm bandpass) and SiV spectral band (715~nm longpass).
(c) Standard confocal microscopy images with 515~nm laser of individual NV centers in SiV center rich Sample A and SiV center absent Sample B, showing the same spatial regions as the measurements in (d), with circled NV centers used for respective measurements in (d).
(d) SPaCE measurement of Sample A ($\tau$ = 5~ms) and Sample B ($\tau$ = 10~ms). The color scale represents the average probability that the measured NV center is in NV$^-$ after the CPG pulse at that location.}
\end{figure}

In this work, we use an individual NV center to study charge dynamics in diamond. These dynamics include optically changing a defect's charge state through the photogeneration of free charge carriers from that defect, as well as the transport and capture of these charge carriers by other defects. 
We define the photogeneration of holes as the optical release of holes from a defect state to the valence band via photoexcitation of electrons from the valence band to the defect. Alternatively, the photogeneration of electrons refers to the photoexcitation of an electron from the defect state into the conduction band.  
In prior works, the average charge state of ensembles of NV centers in diamond has been observed to change due to the capture of charge carriers generated by optical illumination of defects several microns away \cite{Jayakumar2016, Dhomkar2018, Lozovoi2020, Jayakumar2020}. However, the same ensemble of NV centers was used to both produce and capture the charge carriers, leading to a complex interplay of the processes involved. Our use of single-shot charge state readout of an isolated NV center to probe the capture of holes produced by hole photogeneration of a different defect species allows us to isolate the different processes. We use this method (Fig.\ref{fig:SPaCE_measurement}(a)) to investigate the charge dynamics of ensembles of SiV centers and observe exchange of charge carriers between SiV centers and a single NV center.  Furthermore, while  SiV$^{-}$ centers are known to enter a dark state under optical illumination \cite{Dhomkar2018, Nicolas2019}, the charge of this dark state is still debated \cite{Nicolas2019}, alternatively being assigned to SiV$^0$ \cite{Dhomkar2018, Haenens2011, Green2019} or SiV$^{2-}$ \cite{Gali2013, Breeze2020}. The mechanism by which a specific SiV center is returned to the SiV$^{-}$ charge state is also not understood. Our measurements present clear evidence that illumination of an SiV center in the SiV$^{-}$ charge state with 515~nm, 589~nm, or 638~nm light results in hole generation and conversion to the optically dark and spin-less SiV$^{2-}$ state, while illumination of other nearby defects can be used to return it to the SiV$^{-}$ state via hole capture. This new measurement technique provides a unique probe of the microscopic charge dynamics in diamond and a novel tool for realizing new levels of charge control over individual defects. 

Experiments were conducted under ambient conditions using a home-built confocal microscope, with a 1.3~NA oil immersion objective. The laser focus was positioned at x and y using scanning galvanometer mirrors and a 4-$f$ lens configuration. The focus was fixed at $\sim$30~\textmu m below the diamond surface to mitigate surface effects.

\begin{figure}
\includegraphics[width=0.96\textwidth]{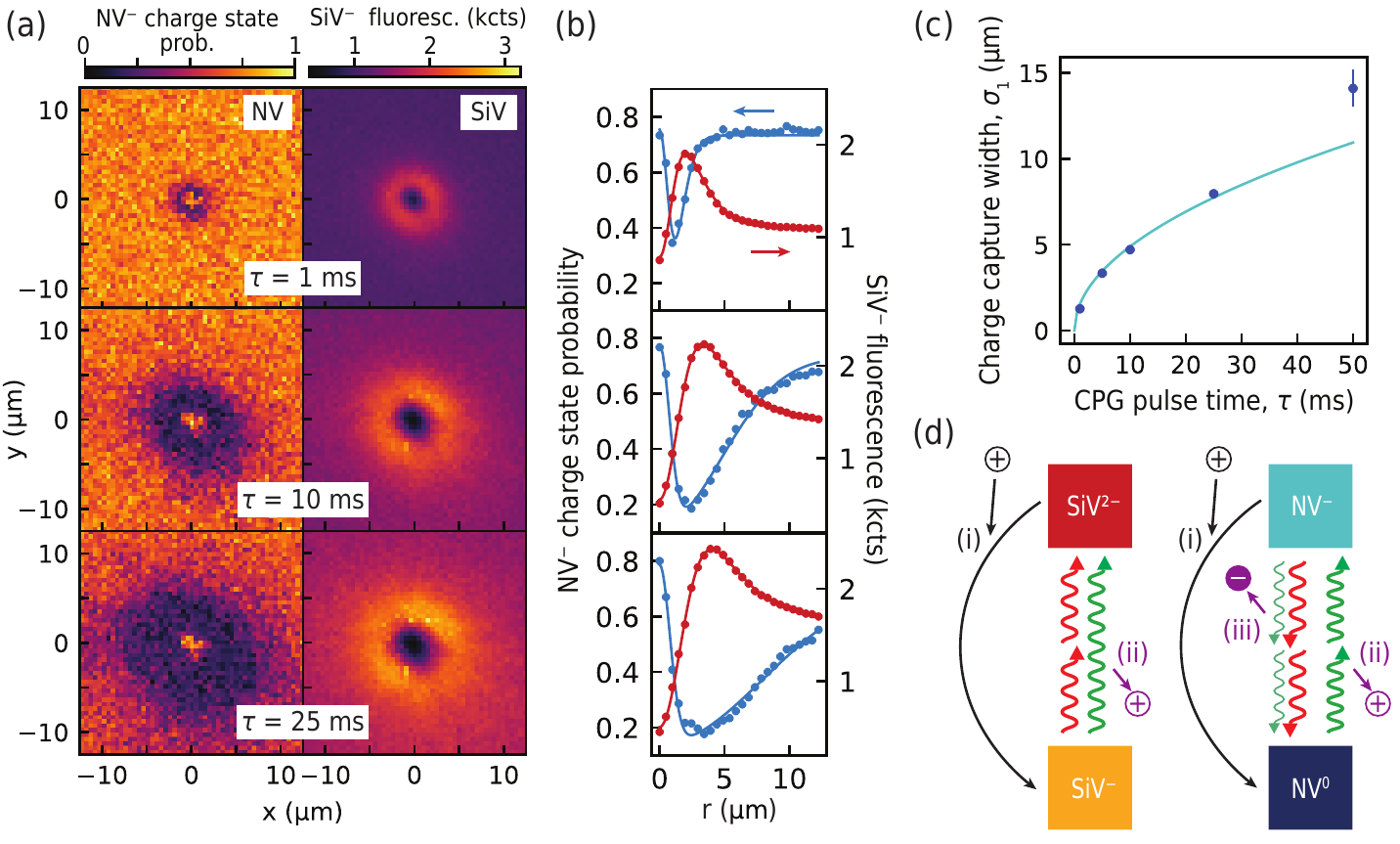}
\caption{\label{fig:vary_tau} SPaCE dependence on CPG pulse length (a) SPaCE measurement for different CPG pulse lengths (additional times in Supplemental Material \cite{SupplementalMaterials}), comparing readout in the NV spectral band (left column) or SiV spectral band (right column). Data in SiV band was taken with 0.4~mW, 40~ms 589~nm readout pulse. The SiV center color scale corresponds to the SiV$^-$ fluorescence counts.
(b) Radial averages comparing the NV (solid blue circles, left axis) and SiV (solid red circles, right axis) measurements of the same CPG pulse length in (a), along with fits to Eq.~\ref{eq:double_gaussian} for the NV (blue line) and SiV (red line) radial averages.
(c) The width, $\sigma_1$, of the hole capture Gaussian curve extracted by  fitting Eq.~\ref{eq:double_gaussian}  as a function of $\tau$, and a fit to an effective hole diffusion process with diffusion coefficient $D_{eff}=1.2$~\textmu m$^2$/ms. We exclude the point at $\tau = 50$ ms from the fit due to saturation of the NV center's charge state at long CPG pulse times (see Supplemental Material \cite{SupplementalMaterials})
(d) Diagram illustrating the charge dynamics observed between relevant charge states of SiV and NV centers: (i) capture of free holes (black arrows), (ii) photogeneration of holes, (iii) photogeneration of electrons. Green and red wavy arrows depict 515 nm and 638 nm illumination, respectively, while one or two arrows depict one- or two-photon photogeneration processes, respectively.}
\end{figure}

Two as-grown chemical-vapor deposition bulk diamond samples were used in our experiments. Sample A, from Diamond Elements, has an NV center concentration of $10^{-3}$~ppb, as well as a high concentration of SiV centers. The density of native NV centers is low enough to resolve and address them individually. The concentration of SiV centers in Sample A was not measured directly, but is at least 10~ppb, as individual SiV defects are not optically resolvable. Sample B, from Element Six, has an NV center concentration of $10^{-5}$~ppb, with no detectable concentration of SiV centers. 
The N defect concentrations of Sample A and B are assumed to be roughly $0.3$~ppb and $3\times10^{-3}$~ppb, respectively, based on the native NV concentration \cite{Edmonds2012}. 
Figure~\ref{fig:SPaCE_measurement}(b) shows the fluorescence spectrum of the NV$^-$ (i) and  SiV$^-$ (ii) charge states. The separation in wavelength of the NV$^-$ and SiV$^-$ emission enables spectral filtering to isolate either the NV$^-$  or SiV$^-$ fluorescence as depicted by the respective blue and red band in Fig.~\ref{fig:vary_tau}(b).

In order to study defect charge conversion and the photogeneration, transport, and capture of charge carriers in diamond we introduce a new technique that makes use of charge-state readout of a single NV center, which we call ``Spatial Photogeneration and Capture of chargE" (SPaCE). The experimental sequence is illustrated in Fig.~\ref{fig:SPaCE_measurement}(a). First, a single spatially resolved NV center at position (0, 0) 
(see Fig.~\ref{fig:SPaCE_measurement}(c)) is prepared by an ``initialization" pulse. In the pulse sequence shown here, the initialization laser pulse is 515~nm (green), which prepares the NV in the NV$^-$ charge state with $\sim 70\%$ probability \cite{SupplementalMaterials, Waldherr2011}. The microscope focus is moved to a new spot (x$_i$, y$_i$)
in the confocal plane (at the same depth below the diamond surface) and a ``charge photogeneration " (CPG) pulse is applied for time $\tau$. The defects at this location may photogenerate holes and electrons, which 
travel out into the diamond and can be captured by the NV center, altering its charge state. 
Finally, a low power 589~nm (yellow) ``readout" laser pulse is used to measure the resulting charge state of the NV at (0,0) \cite{Shields2015} (see Supplemental Materials for more details  \cite{SupplementalMaterials}). This sequence is repeated for different (x$_i$, y$_i$) positions of the CPG pulse and the entire measurement sequence is repeated multiple times to produce a spatial map of the average charge state probability of the NV center as a function of CPG pulse location.

Figure~\ref{fig:SPaCE_measurement}(d) shows the outcome of SPaCE experiments performed on Sample A and Sample B, measured on the respective single NV centers circled in (c). 
There are three distinctive features in the measurement from Sample A. First, the bright region in the center corresponds to the CPG pulse directly illuminating the NV center, leaving it in the NV$^-$ state with a $\sim 70\%$ probability. The size of this central region is related to the CPG pulse beam waist, duration, and intensity. Second, there is a bright area around the edges of the plot corresponding to CPG pulses positioned at radial distances greater than 5~microns from the NV center, which for the 5~ms pulse duration shown here leaves the NV center charge state unaffected.

Lastly there is a dark ring centered on (0,0)~microns, which corresponds to application of the CPG pulse at a radial distance between $\sim$ 1~\textmu m and 5~\textmu m from the NV center. With the 5~ms CPG pulse positioned within this range, the NV center had a high probability to be converted from NV$^-$ to NV$^0$ at some point during the pulse.

For comparison, the same SPaCE measurement on an NV center in Sample B does not produce a dark ring, even at longer CPG pulse times (up to 100~ms). This implies the NV$^{0}$ ring's presence in Sample A and absence in Sample B is due to a difference between the two samples (the same behaviors were also observed with many other NVs in both samples). While there are differences in NV center and N defect concentrations between the two samples, we argue that these differences cannot account for the observations. A comparison of Sample B's confocal scan in Fig.~\ref{fig:SPaCE_measurement}(c) and its SPaCE measurement in Fig. 1(d) (which represent the same spatial area) reveals that the central NV center remains bright when the CPG pulse is positioned over other nearby NV centers. We conclude that for the pulse durations, powers, and defect densities used in this work, charge carriers generated by illumination of surrounding isolated NV centers do not play a role in the observed charge dynamics. Furthermore, substitutional N defects do not produce holes under 515~nm illumination \cite{Isberg2006, Jayakumar2016,Dhomkar2018}. While 515~nm illumination can photoionize N$^0$ to N$^+$, releasing electrons into the conduction band, this cannot explain the conversion of NV$^{-}$ to NV$^{0}$. We therefore attribute the dark ring observed in the SPaCE measurements in Sample A to charge capture of holes released into the valence band by optical charge conversion of SiV centers during the CPG pulse, which we know to be present in Sample A and not in Sample B. 

The attribution of the charge state conversion to holes photogenerated by charge conversion of SiV centers can be confirmed by taking advantage of the resolved defect spectra shown in Fig.~\ref{fig:SPaCE_measurement}(b). The outcome of identical SPaCE measurement sequences for an NV center and an ensemble of SiV centers are shown in the left and right column of Fig.~\ref{fig:vary_tau}(a), respectively, for varying CPG pulse lengths, $\tau$ (see Supplemental Materials for additional measurements \cite{SupplementalMaterials}). As opposed to the single NV center measurements, the density of the SiV defects is too high to isolate individual SiV centers and their concentration can only be estimated, so a single-shot charge state probability cannot be assigned to the SiV centers. Instead, these measurements extract the relative SiV$^-$ fluorescence following the SPaCE sequence from an ensemble of readout SiVs at the center (spatial position (0,0) microns, same as the NV center), with the number of SiV centers contributing to the measurement determined by the confocal volume and the local SiV density. 

As shown in Fig.~\ref{fig:vary_tau}(a), SPaCE measurements performed with readout in the SiV spectral band exhibit similar but inverted features relative to the NV SPaCE measurements. The dark region in the center, which corresponds to applying the CPG pulse directly to the central readout SiV centers, indicates that a millisecond (or longer) 515~nm laser pulse converts SiV centers to a dark state, consistent with prior reports \cite{Dhomkar2018}. When the laser is pulsed at larger radial distances around the edges of the plots, the fluorescence from the central SiV centers is also low. This corresponds to the SiV charge state after the previous measurement and initialization pulse, which together prepare them mainly in the dark state. Finally, in contrast to the dark ring in the NV measurements, the SiV centers exhibit a bright ring, which corresponds to the CPG pulse positioned a few microns away causing the central readout SiV centers to convert from their dark state to the bright SiV$^-$ charge state. This strongly suggests that just as the NV center captures holes, the SiV centers also capture holes photogenerated by charge conversion of surrounding SiV centers, resulting in the bright ring. Based on the combined observations that direct illumination with green light converts the SiV center into the dark state, the conversion of NV$^{-}$ to NV$^{0}$ during SiV photo-conversion, and the strong anti-correlation between NV and SiV charge states, we conclude that the observed SiV dark state must be SiV$^{2-}$. This is further supported by the lack of any fluorescence signal from SiV$^0$ in this sample (see Supplemental Material \cite{SupplementalMaterials}).

Figure~\ref{fig:vary_tau}(b) shows the radial averages \cite{SupplementalMaterials} of the SPaCE measurements as a function of distance from the center. The data is well described by the difference of two Gaussian functions:
\begin{equation}\label{eq:double_gaussian}
    G(r)=C - A_1^2 e^{-r^2/(2 \sigma_1^2)} + A_2^2 e^{-r^2/(2 \sigma_2^2)},
\end{equation}
where $C, A_1, A_2, \sigma_1$ and $\sigma_2$ are fit parameters and are allowed to vary for different CPG pulse times, $\tau$. The two Gaussian functions in Eq.~\ref{eq:double_gaussian} correspond to two competing processes:  capture of free holes and direct photoconversion by the laser. For the NV measurements the width of the negative Gaussian, $\sigma_1$, corresponds to the hole capture process. This parameter is extracted from the NV data fits as a function of $\tau$ (see Supplemental Material for additional measurements and details \cite{SupplementalMaterials}), and is shown in Figure~\ref{fig:vary_tau}(c). For CPG pulse times less than or equal to 25 ms, $\sigma_1$ is well described by  
$\sigma_1 = (2 D_{eff})^{1/2} \tau^{1/2}$, where $D_{eff}=1.2$~\textmu m$^2$/ms is an effective hole diffusion constant in the presence of SiV centers and N defects for the sample and laser power used here. 

In order to better understand the origin of the effective diffusion constant, we applied a model following Refs.~\citenum{Dhomkar2018} and \citenum{Jayakumar2020} that consists of a system of equations describing the diffusion of photogenerated charge carriers (holes and electrons) and dynamics of carrier capture by the defects present in the sample (N$^0$/N$^+$, NV$^-$/NV$^0$, SiV$^{2-}$/SiV$^-$; see Fig.~\ref{fig:vary_tau}(d) and Supplemental Materials \cite{SupplementalMaterials}). Using the cylindrical symmetry approximation employed in Refs.~\citenum{Dhomkar2018} and \citenum{Jayakumar2020}, we could not obtain a good simultaneous fit to all the experimental radially-averaged SPaCE data using a single set of adjustable parameters, even though the model could qualitatively reproduce the shape of the data (see the Supplemental Material \cite{SupplementalMaterials}). Our analysis found the simulations strongly depend on the SiV$^{2-}$ hole capture rate, which varied over an order of magnitude between fits of different CPG pulse times. We conjecture that the failure of this model to quantitatively reproduce the data with a single set of free rate parameters suggests that the effective diffusion constant arises in part from charge dynamics taking place outside of the focal plane, which are not captured by the cylindrically symmetric model. We leave simulations in three dimensions for future work.

Figure~\ref{fig:vary_tau}(d) depicts the three relevant charge carrier capture and charge carrier photogeneration processes observed in this work. Hole photogeneration through defect charge state conversion can occur for NV centers (NV$^0$ to NV$^-$) under illumination with 515 nm light \cite{SupplementalMaterials, Waldherr2011, Aslam2013} and for SiV centers (SiV$^-$ to SiV$^{2-}$) under illumination with both 515 nm light and 638 nm light (see Supplemental Materials \cite{SupplementalMaterials}). Electron release through the photoionization of NV$^-$ to NV$^0$ occurs under 515 nm light (at a slower rate than the opposite process) and 638 nm light, but does not occur for the SiV center for optical wavelengths. Finally, we find that both NV and SiV centers capture free holes (black arrows), but do not capture free electrons at significant rates. Hole capture is therefore the only way to convert from SiV$^{2-}$ to SiV$^-$ based on our observations, although photoionization of SiV$^{2-}$ may be possible with UV wavelengths \cite{Breeze2020}. This indicates that in applications involving individual SiV$^-$ centers  \cite{Bhaskar2020}, returning to the SiV$^-$ charge state from the dark state can be accomplished by shining 515 nm light on other nearby defects, and not necessarily on the SiV center itself. 

\begin{figure}
\vspace{-0.3cm}
\includegraphics[width=0.48\textwidth]{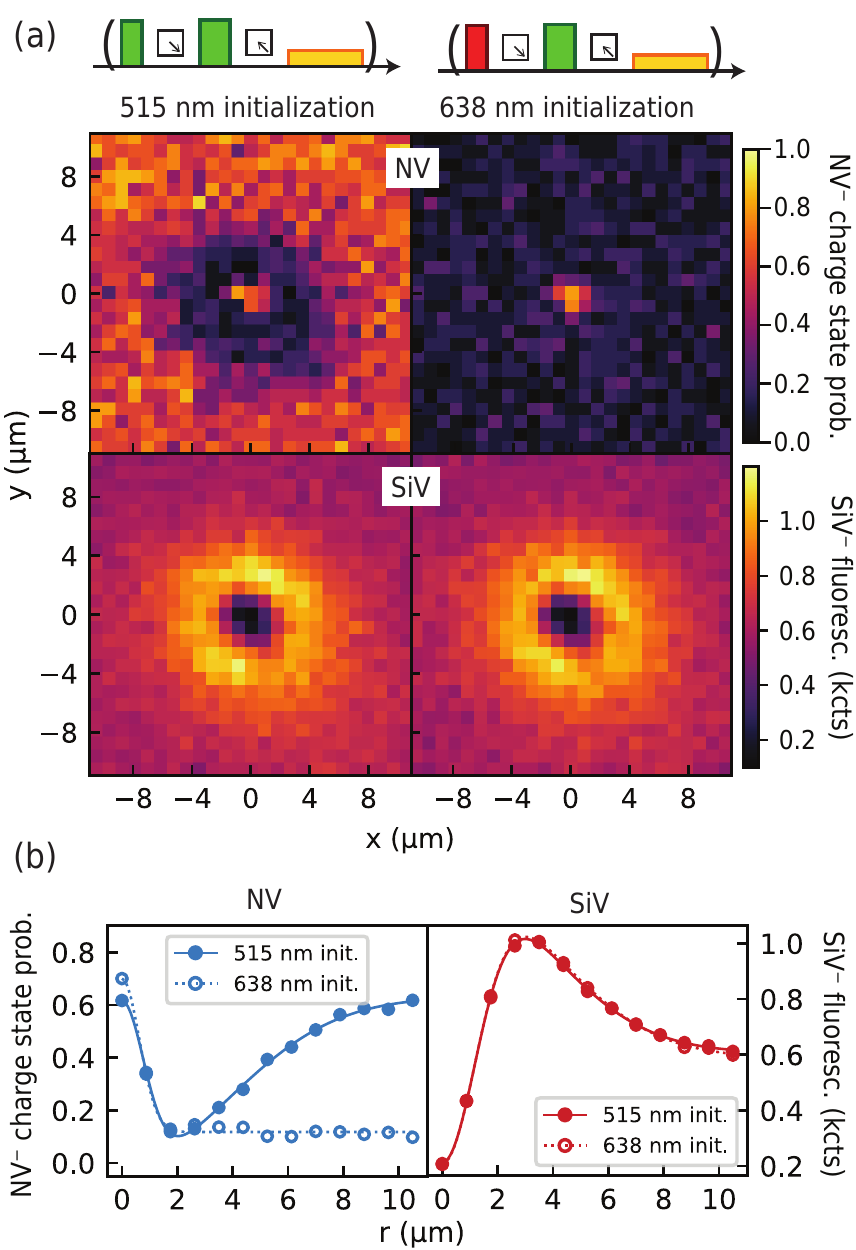}
\caption{\label{fig:NV_init} SPaCE measurements with the NV center initialized in different charge states. (a) SPaCE measurements with NV$^-$ initialization (515~nm, 2~mW, 1~\textmu s) is shown in the left column and with NV$^0$ initialization (638~nm, 8~mW, 1~\textmu s) in the right column. Dynamics in the NV and SiV spectral band are compared and shown in the top and bottom rows, respectively. For all measurements, $\tau = 10$~ms.
(b) The corresponding radial average points of the NV center (blue circles) and SiV center (red circles) are shown with fitted lines to Eq.~\ref{eq:double_gaussian} for 515 nm initialization (filled circles, solid lines) and 638 nm initialization (open circles, dotted lines). 
 }
\end{figure}

In Fig.~\ref{fig:NV_init}(a), SPaCE measurements are shown with a 638~nm (red) initialization pulse which prepares the central NV in the NV$^0$ state \cite{Aslam2013} (top right), in contrast to the 515~nm initialization pulse used in Figs.~\ref{fig:SPaCE_measurement} and \ref{fig:vary_tau} that prepare the central NV center predominantly in the NV$^{-}$ state (top left). For 638~nm initialization, only a bright central region is observed. This corresponds to the CPG pulse directly illuminating the central NV center and directly converting to $70\%$ NV$^-$ from the NV$^0$ state. The fact that outside this region the NV center remains in the NV$^0$ state indicates that on these time scales the NV center does not capture electrons generated from other photoionized defects, such as N defects. This is in agreement with previous reports that the NV$^0$ electron-capture cross-section is consistent with zero \cite{Dhomkar2018}. Also shown are the measurements using the SiV spectral band for 638~nm initialization (bottom plots of Fig.~\ref{fig:NV_init}(a) and right plot of (b)). The SiV charge state dynamics are unchanged by the wavelength of the initialization pulse. 

\begin{figure}
\vspace{-0.3cm}
\includegraphics[width=0.48\textwidth]{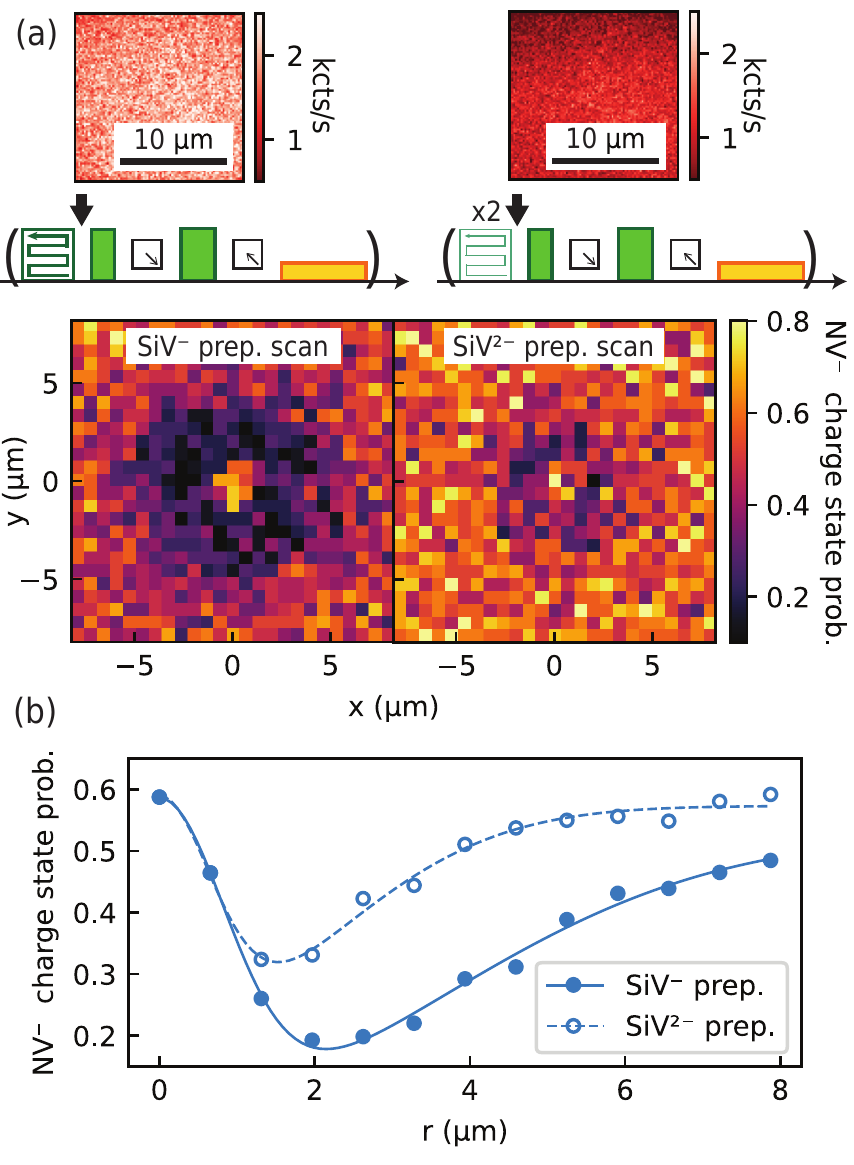}
\caption{\label{fig:SiV_prep} Optical engineering of the surrounding defect charge state environment. (a) SiV centers in the surrounding area are prepared prior to each SPaCE measurement sequence by performing a 515~nm laser raster scan, represented by the zig-zag block in pulse train. The SiV$^-$ charge state (left) is prepared with a 1~mW preparation scan and 10~ms dwell time at each point, and SiV$^{2-}$ is prepared with 20~\textmu W preparation scan and same dwell time, performed twice. Representative SiV center fluorescence after the preparation scans are shown in confocal images at the top of the figure, taken with 589~nm (3~\textmu W, 20~ms) readout in the SiV band. The corresponding SPaCE measurements on a single NV center for the two SiV preparations ($\tau$ = 10~ms) are shown in the bottom panels.
(b) Radial average of the NV$^-$ charge state probability and fitted lines to Eq.~\ref{eq:double_gaussian}, with SiV$^-$ (solid circles, solid line) and SiV$^{2-}$ (open circles, dashed line) preparation.  
 }
\end{figure}

To demonstrate that the impacts of charge capture can be mitigated by control over the surrounding charge environment, SPaCE measurements were performed after preparing the surrounding SiV centers in a specific charge state before each initialization pulse. As shown in Fig.~\ref{fig:SiV_prep}(a), the measurement area was raster scanned to prepare a majority of the SiV centers in either the SiV$^-$ or SiV$^{2-}$ charge state following the protocol developed in Ref.~\citenum{Dhomkar2018}. Confocal scans over the same area as the SPaCE measurements show a homogeneous SiV charge state after the raster scan, with the count rate in the SiV band after SiV$^-$ preparation approximately double those after SiV$^{2-}$ preparation.

The corresponding SPaCE measurements performed on the NV center are shown in Fig.~\ref{fig:SiV_prep}(a). The two measurement sequences are identical, with the only difference between the two data sets being the SiV center preparation raster scans performed prior to each SPaCE measurement sequence. As the initialization pulse is applied after the raster scan, the NV center charge state is not affected by the SiV center preparation, but the subsequent charge dynamics are. When the surrounding SiV centers are prepared primarily in the SiV$^{2-}$ charge state (right), the probability that an NV captures a hole and is converted to NV$^{0}$ is reduced significantly compared to when the surrounding SiV centers are prepared in the SiV$^{-}$ charge state (left). Figure~\ref{fig:SiV_prep}(b) shows the radial averages of the measurements in (a), along with fits to Eq.~\ref{eq:double_gaussian}. The width of the charge capture Gaussian curve, \cite{SupplementalMaterials} $\sigma_1$, was reduced by a factor of 1.6 by preparing the surrounding SiV centers in the SiV$^{2-}$ state.

In conclusion, we have presented a new measurement technique that makes use of single NV centers to probe photogeneration of holes from surrounding defects, hole transport, and hole capture. Using this technique, hole exchange between SiV centers and a single NV center over micron distances was observed. These observations show that the dark state of the SiV center under optical illumination is the SiV$^{2-}$ charge state, and that the primary mechanism for its return to the bright SiV$^{-}$ state is through charge capture of holes photogenerated by charge conversion of other surrounding defects. We have also demonstrated that the impact of hole capture on a single NV center can be partially mitigated by engineering the surrounding defect charge state environment. The SPaCE measurement technique demonstrated here can be readily applied to any diamond sample containing spatially resolved NV centers, and is adaptable to any color center where the charge state can be optically measured, enabling the future study of the charge dynamics of a broad range of defects. Finally, the observation of charge conversion of single NV centers due to charge exchange with defects many microns away highlights the importance of considering charge dynamics in experiments with arrays of color centers, and points the way to new methods of color center charge state engineering and control. 

\emph{Author's note:} During preparation of this manuscript we became aware of complementary work in which charge transport and capture between two individual nitrogen vacancy centers in diamond was observed \cite{LozovoiConf2021, LozovoiSubmitted}.

\section{Acknowledgements}
\vspace{0.5cm}
\noindent The authors thank Nathalie de Leon for enlightening discussions, helpful insights, and comments on the manuscript. Experimental work, data analysis, and theoretical efforts conducted at UW--Madison and Lawrence Livermore National Laboratory were supported by the U.S. Department of Energy (DOE), Office of Science, Basic Energy Sciences (BES) under Award \#DE-SC0020313. Part of this work by V.L. was performed under the auspices of the U.S. Department of Energy at Lawrence Livermore National Laboratory under Contract DE-AC52-07NA27344. Theoretical contributions conducted at Pontificia Universidad Cat\'olica de Chile by J.~R.~M.~ and A.~N.~ were supported by ANID Fondecyt 1180673 and ANID PIA ACT192023. A.~N.~ and H.~T.~D. acknowledge financial support from Universidad Mayor through the Postdoctoral Fellowship. A.~G.~acknowledges support from the Department of Defense through the National Defense Science and Engineering Graduate Fellowship (NDSEG) program.

\bibliography{MainReferences}

\end{document}



\section{Experimental methods and additional data}

\subsection{Second-order photon correlation measurements}

\begin{figure}
\vspace{-0.3cm}
\includegraphics[width=0.9\textwidth]{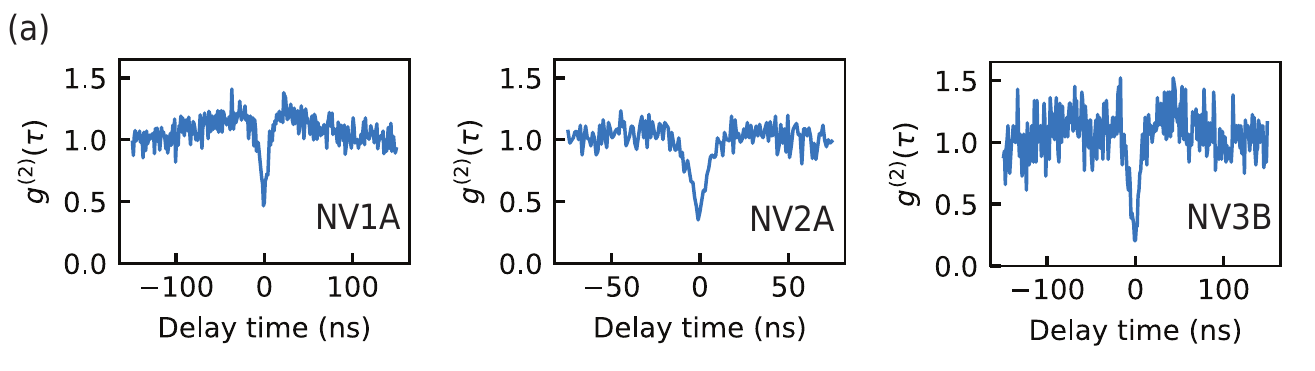}
\caption{\label{fig:g2} Second-order photon correlation function $(g^{(2)}(\tau)$) measurements of the three NVs presented in this paper, without background subtraction. Integration time was 9 min for NV1A, 10 min for NV2A, and 12 min for NV3B. For NV1A $g^{(2)}(\tau=0)=0.49$, for NV2A $g^{(2)}(0)=0.42$, for NV3B $g^{(2)}(0)=0.20$.
}
\end{figure}

Data from measurements performed on three individual nitrogen-vacancy (NV) centers are presented in the main text. NV1A is in Sample A and is used for measurements presented in Figs. 1-3 of the main text and in supplemental figures. NV2A is also in Sample A and used for Fig. 4 of the main text. NV3B is in Sample B and used only for Fig. 1 of the main text. Figure~\ref{fig:g2} shows the second-order photon correlation function $g^{(2)}(\tau)$, without background subtraction, for each of these NV centers. All three NV centers exhibit a $g^{(2)}(0) < 0.5$.

\subsection{Measurements of other NV centers in Sample A}

\begin{figure}
\vspace{-0.3cm}
\includegraphics[width=0.4\textwidth]{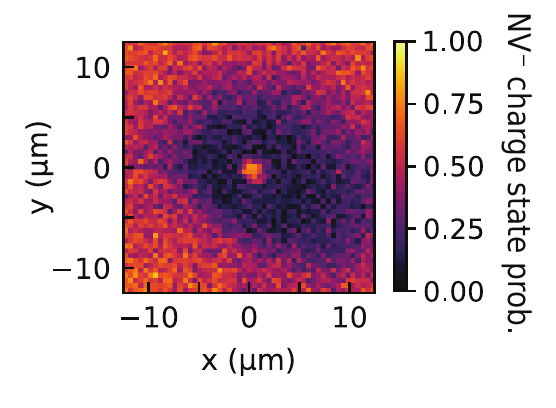}
\caption{\label{fig:space_assymetry} Spatial Photogeneration and Capture of chargE (SPaCE) measurement with an additional NV center (not discussed in this paper) exhibiting asymmetry in NV$^0$ dark ring (measurement parameters same as those in Fig.~\ref{fig:additional_taus}, $\tau$ = 25~ms).
}
\end{figure}

Spatial Photogeneration and Capture of chargE (SPaCE) measurements were also performed with $\sim$10 other NV centers in Sample A that were not included in this paper (data available upon request). All such measurements showed qualitatively similar behavior, although some NV centers exhibited asymmetries in the shapes of the NV$^0$ charge state rings (see Fig.~\ref{fig:space_assymetry}). The nature of these asymmetries is unclear, but may be due to variations in the local SiV defect density or the presence of other charged defects in the sample \cite{Lozovoi2020SpaceCharge}.

\subsection{SPaCE measurement protocol}
As stated in the main text, the SPaCE measurements on both a single NV center and ensembles of SiV centers presented in the paper were averaged over multiple measurements. Plots shown were averaged over 20 to 30 measurements. For each individual measurement the order of charge photogeneration (CPG) pulse spatial locations was randomized to mitigate measurement-to-measurement temporal effects.

\subsection{Radial average calculation}
\begin{figure}
\vspace{-0.3cm}
\includegraphics[width=1\textwidth]{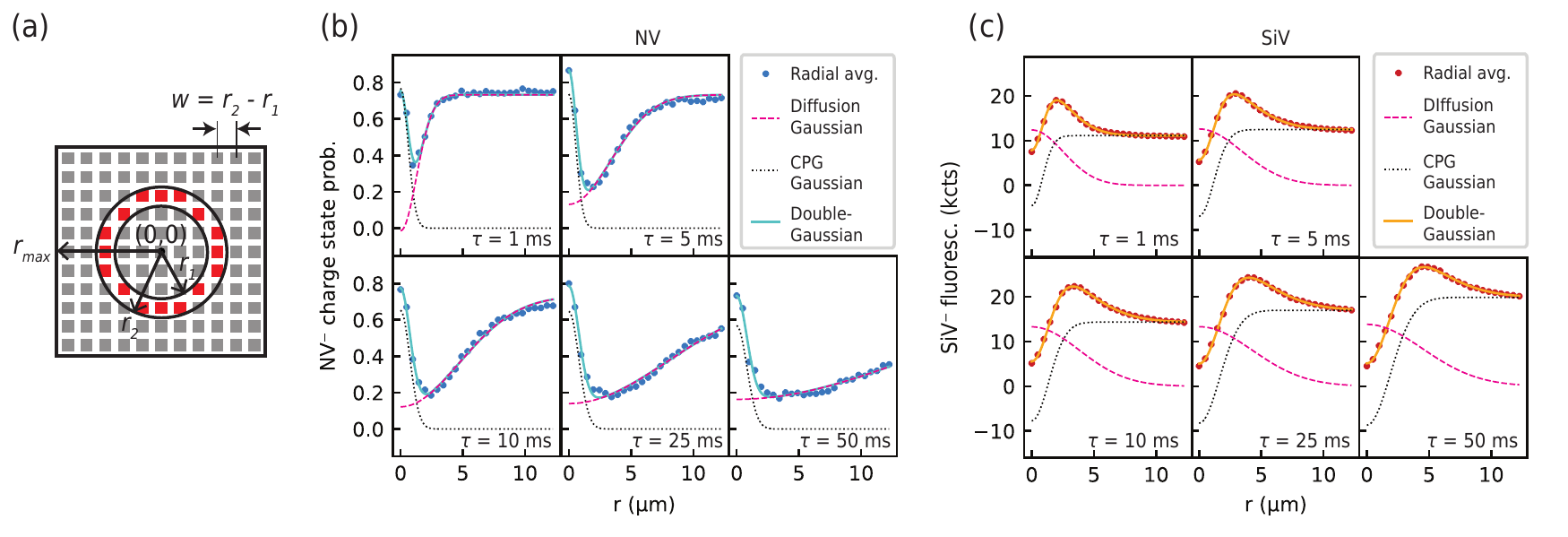}
\caption{\label{fig:rad_avg} Radial averages of SPaCE measurements. (a) Calculation of radial average of SPaCE measurement (represented by grid of pixels). Pixels that share similar radial distances from the center (i.e. the red squares), determined by an annulus centered about (0,0) of width $w$ determined by the pixel size, are averaged together. 
(b) Radial averaged data from NV measurements of various CPG pulse times $\tau$ (blue dots). The combination of two Gaussian curves (black dotted curve and magenta dashed curve) are fit to data, the sum is represented by light blue curve. 
(c) Radial averaged data from SiV measurements of various CPG pulse times $\tau$ (red dots). The combination of two Gaussian curves (black dotted curve and magenta dashed curve) are fit to data, the sum is represented by orange curve. 
}
\end{figure}

To obtain the radial averages shown in Figs.~2-4 of the main text and throughout the supplement, an average is taken over the CPG locations that share a similar radial distance from the center. Figure~\ref{fig:rad_avg}(a) illustrates this process with an array of pixels (representing a SPaCE measurement) and a representative selection of the pixels (red squares) whose radial distance from (0, 0), $r$, falls within $r_1 \leq r < r_2$. An annulus can be used to visualize this selection of points, with inner radius $r_1$ and outer radius $r_2$. These red points, whose centers are contained within the annulus, are averaged together, and this average is considered as the ``radial average" for a radial position equal to $r_1$. This selection and averaging is performed for concentric annuli, each with width $w=r_2-r_1$, where $w$ was selected to be the pixel size for that SPaCE measurement, such that every pixel is exclusively binned into one annulus. This is repeated up until the largest annulus, whose outer radius $r_2 = r_{max}$.

The radial averages are accompanied by fits to  Eq. 1 of the main text. Figure~\ref{fig:rad_avg}(b) and (c) show the radial average data (blue and red circles, respectively) of measurements for varying CPG pulse times $\tau$ (see Fig.~\ref{fig:additional_taus}), along with fits to Eq. 1 (light blue and yellow curves, respectively). Also shown are the individual positive and negative Gaussian curves that, when summed, make up Eq. 1, and are plotted as individual Gaussian functions using the fit values $C, A_1, \sigma_1$ and $A_2, \sigma_2$ (and no vertical offset), respectively. As mentioned in the main text, the $\tau = 50$~ms point is excluded in the diffusion fit of Fig. 2(c) of the main text due to saturation of the single NV charge state in NV$^0$, which results in a flattened, non-Gaussian profile, and because the full extent of the non-saturated portion of the Gaussian is not within the measurement's spatial range.
For all NV  fits to Eq. 1, the vertical offset, $C$, is constrained to a value of $C=0.73$. The vertical offset is a free parameter for the SiV radial average fits. 

\section{Additional data}

\subsection{Additional CPG pulse measurements}
\begin{figure}
\vspace{-0.3cm}
\includegraphics[width=0.5\textwidth]{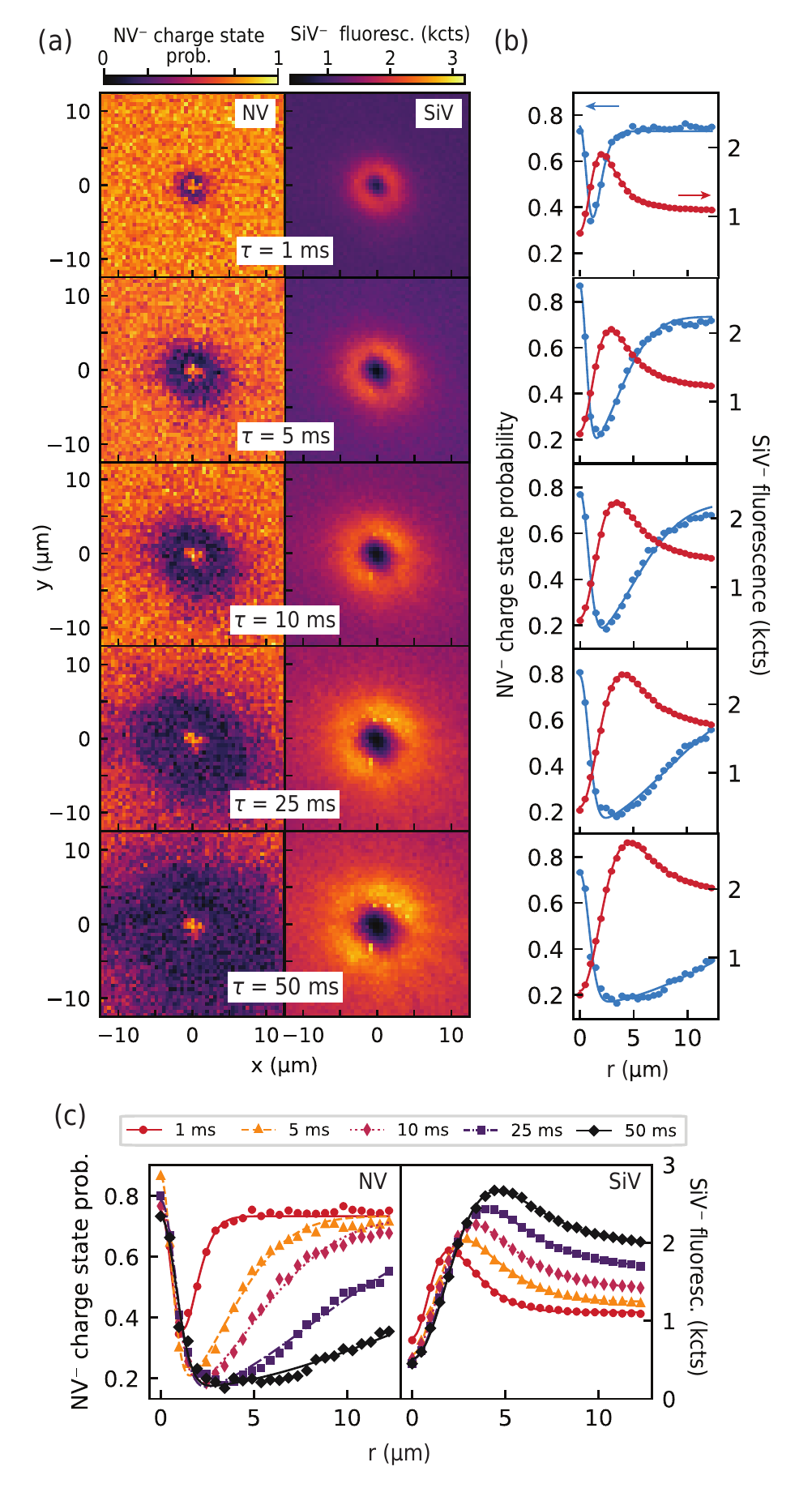}
\caption{\label{fig:additional_taus} SPaCE dependence on CPG pulse length. (a) SPaCE measurement for five different CPG pulse lengths ($\tau = 50$~ms was not shown in main text, for clarity), comparing readout in the NV spectral band (left column) or SiV spectral band (right column). Data was taken with same measurement parameters as described in main text.
(b) Radial averages comparing the NV (solid blue circles, left axis) and SiV (solid red circles, right axis) measurements of the same CPG pulse length in (a), along with fits to Eq. 1 of the main text for the NV (blue line) and SiV (red line) radial averages.
(c) Radial averages comparing measurements of various CPG pulse lengths in (a) and (b) for either readout in NV or SiV spectral band.
}
\end{figure}

SPaCE measurements with CPG pulse lengths $\tau=5$~ms and $\tau=\{1, 10, 25\}$~ms are presented in Figs. 1(d) and 2(a) of the main text, respectively. We collect all of these measurements, including an additional measurement at $\tau=50$~ms (not shown in the main text for clarity) in extended Fig.~\ref{fig:additional_taus}(a). The corresponding radial average curves are shown in Fig.~\ref{fig:additional_taus}(b), and are collectively plotted in (c).

\subsection{Wavelength dependent charge photogeneration}\label{red_photoionization}
\begin{figure}
\vspace{-0.3cm}
\includegraphics[width=0.4\textwidth]{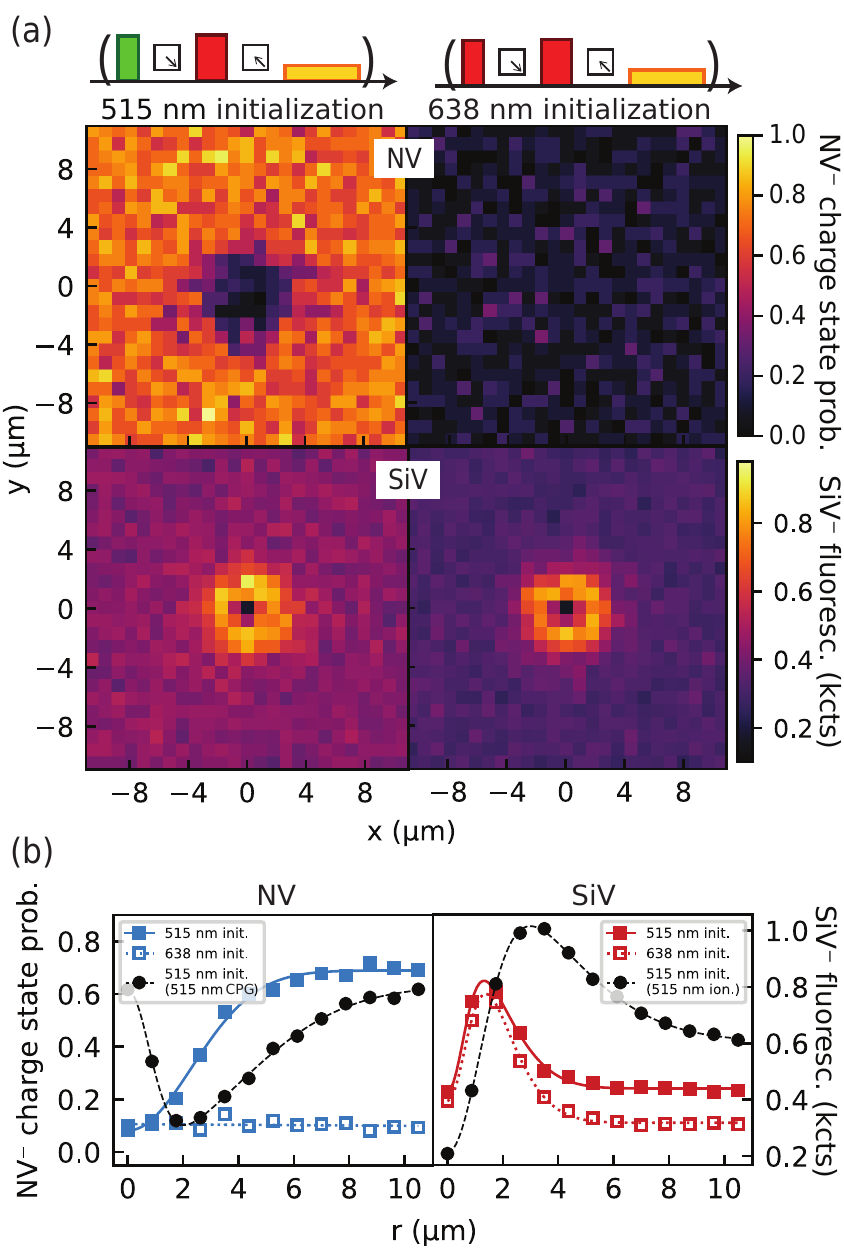}
\caption{\label{fig:red_ionize} SPaCE measurements with 638 nm CPG pulse. (a) SPaCE measurements with readout in NV and SiV spectral band, similar to measurements in Fig. 3 of main text, but with a 638~nm (red) CPG pulse (8 mW, 10 ms). 
(b) The corresponding radial average points of the NV center (left) and SiV center (right) are shown with fitted lines to Eq. 1 for 515 nm initialization (filled squares, solid lines) and 638 nm initialization (open squares, dotted lines). Additionally shown are radial averages (with fits to Eq. 1) of measurements with 515 nm initialization and 515 nm CPG pulse (black circles/dashed line). All measurements were performed with $\tau$ = 10~ms.
}
\end{figure}

In addition to the SPaCE measurements with a 515 nm CPG pulse presented in the main text, SPaCE measurements with a 638 nm CPG pulse were also performed.  Figure~\ref{fig:red_ionize}(a) shows SPaCE measurements with this red CPG pulse performed with both 638 nm (red pulse) and 515 nm (green pulse) initialization pulses, indicated by the pulse sequences. The radial averages of these measurements are shown in Fig.~\ref{fig:red_ionize}(b), along with radial averages from a measurement with 515 nm initialization and CPG, with fits to either Eq. 1 (NV 515 nm init. 515 nm CPG, and all SiV data), a single Gaussian (NV 515 nm init.), or a constant (NV 638 nm init.) As expected, for the NV center measurements (top), in contrast to the main text, the central bright feature is replaced with a dark feature, as the NV center is both photoionized from NV$^-$ to NV$^0$ by the 638 nm CPG pulse laser light and captures holes photogenerated from surrounding defects (see below).  

The bottom measurements in Fig.~\ref{fig:red_ionize}(a) show SPaCE measurements in the SiV spectral band, and show similar features as with the 515 nm CPG pulse from the main text, however, the inner and outer radii of the bright ring are both $\sim$ 2 times smaller than the respective radii of the bright ring with 515 nm CPG with the same pulse time $\tau$ (see Gaussian fits in Fig.~\ref{fig:red_ionize}(b)). Specifically, the smaller and sharper inner radius suggest that the SiV charge state photoexcitation process with 638 nm light is a two photon process. Measurements of the power dependence of the SiV hole photogeneration rates also show a linear dependence with 515 nm and a quadratic dependence with 638 nm (see the section on SiV charge photogeneration rates below). 
Furthermore, the size of the SiV bright ring matches the size of the NV dark feature, indicating that the NV center captures holes photogenerated from the SiV centers, indicating that 638 nm light also photoexcites SiV$^-$ to SiV$^{2-}$.

\subsection{Absence of SiV\(^0\) fluorescence}
\begin{figure}
\vspace{-0.3cm}
\includegraphics[width=0.96\textwidth]{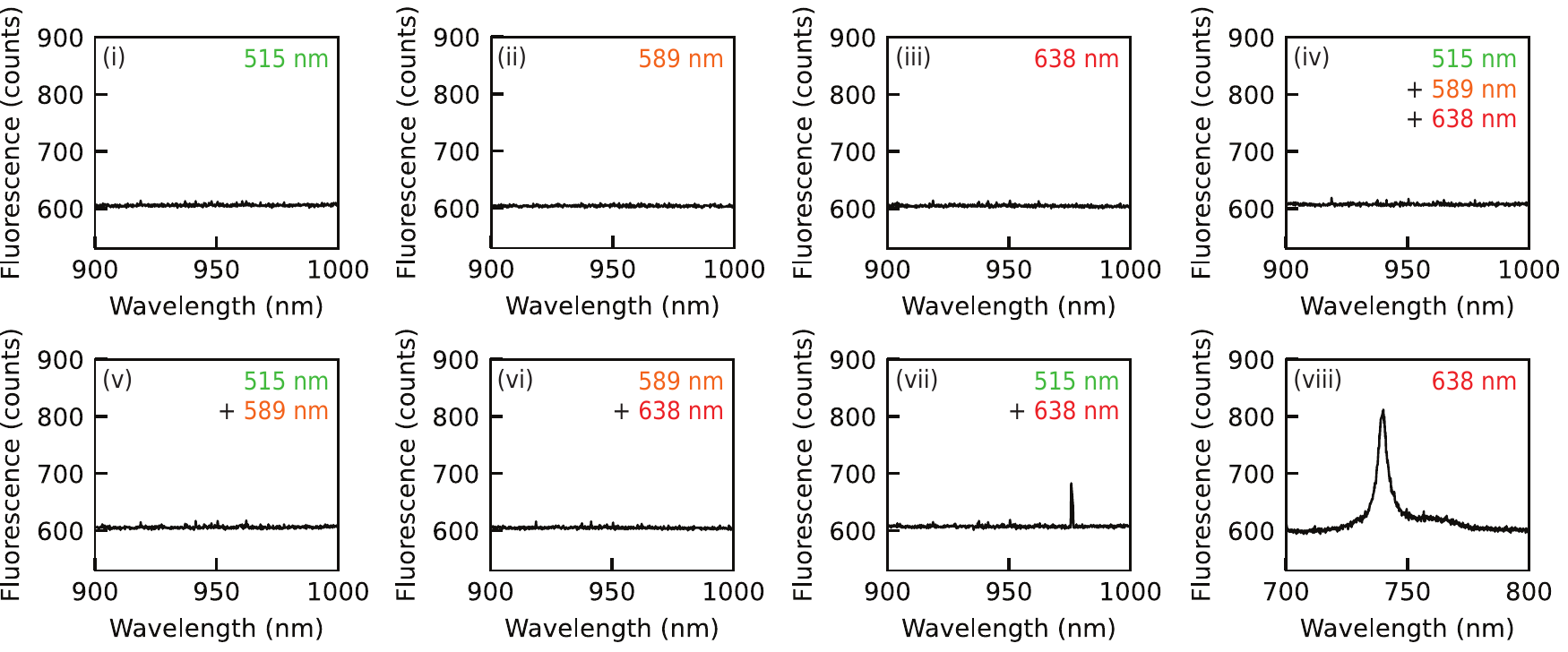}
\caption{\label{fig:siv0_spectra} Fluorescence spectra of Sample A about the SiV$^0$ zero phonon line (ZPL) under various illumination wavelengths, both individually and simultaneously. All laser powers were set to maximum power for our system:  515 nm at 20 mW, 589 nm at 0.38 mW, and 638 nm at 20 mW. For (i) - (vii), 360 s integration time was used. Plot (viii) shows the spectrum of SiV$^-$ ZPL in Sample A, taken with 20 mW, 638 nm laser for 1~s integration time.
}
\end{figure}

Figure~\ref{fig:siv0_spectra}(i - vii) shows measured fluorescence spectra of Sample A around the SiV$^0$ zero phonon line (ZPL) wavelength of 946 nm \cite{Haenens2011} under laser excitation by 515 nm, 589 nm, and 638 nm laser pulses individually and in combination. The maximum achievable laser powers in our microscope (up to 20~mW) were used. Our optical components and spectrometer have an optical efficiency of $\sim 2\%$ at 946 nm based on the manufacturer specifications. No emission from the SiV$^0$ charge state \cite{Haenens2011} was observed.
These spectra around 946 nm are compared to spectrum (viii) of Fig.~\ref{fig:siv0_spectra}, which shows the ZPL of SiV$^-$ under red illumination for 1~s integration. The absence of any signal at 946 nm after 360~s of integration adds an additional piece of evidence that SiV$^0$ is not the ``dark state'' we observe under optical illumination, and agrees with our assignment of SiV$^{2-}$ as the dark charge state.

\section{Single-shot NV charge state readout}

\begin{figure}
\vspace{-0.3cm}
\includegraphics[width=0.6\textwidth]{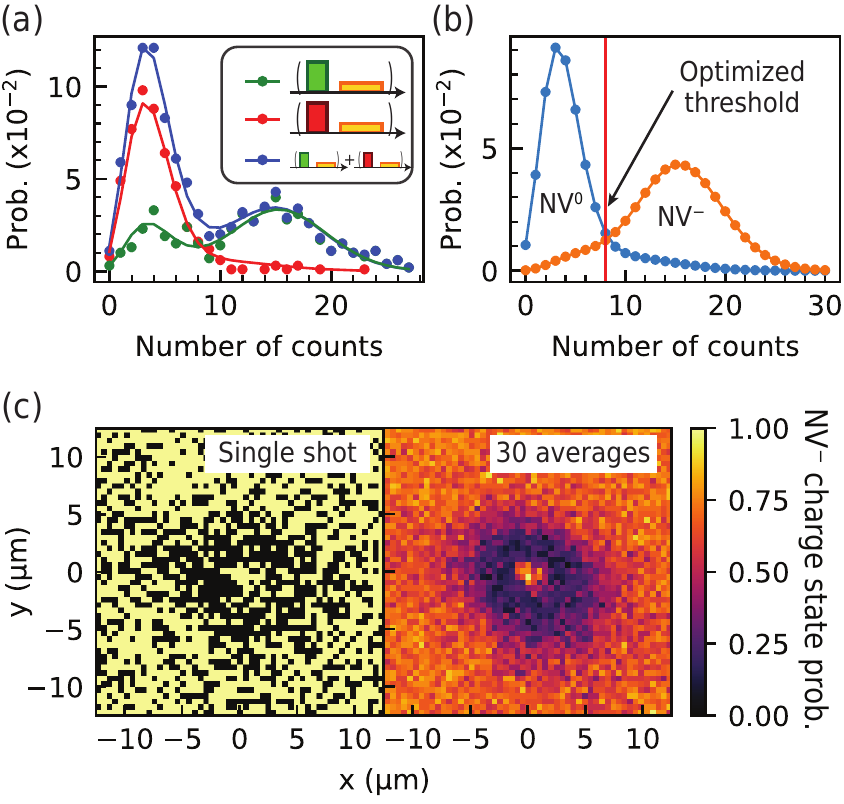}
    \caption{Single-shot charge state readout of an NV center. (a) Photon statistics from a 589 nm readout window (3~uW, 100~ms) after 515 nm (green points) or 638 nm (red points) initialization, 500 measurements each. Blue points are the sum of the two measurements. Green, red, and blue lines are curves following model presented by \citeauthor{Shields2015} \cite{Shields2015}
    (b) The modeled photon number distributions of NV$^-$ and NV$^0$, 
    based off of charge photogeneration and fluorescence rates extracted from (a) and assuming equal probability of an NV center being in either state. The optimized threshold determines the fidelity ($90\%$ for this measurement) of single-shot charge state identification between the two charge states. (c) SPaCE measurement from Fig.~\ref{fig:additional_taus} ($\tau=10$~ms). A single-shot measurement is shown on the left in which charge-state readout is applied: each point is either assigned 0 (NV$^0$) if the counts during readout are below the optimized threshold or 1 (NV$^-$) if the counts are at or above the optimized threshold. The average of 30 singe-shot measurements is shown on the right.}
    \label{fig:sup}
\end{figure}

In the SPaCE measurements presented in this paper, single-shot charge state readout is performed to assign the NV center's state after each measurement (assumed to be either NV$^-$ or NV$^0$). This readout is performed with illumination with a 589~nm laser at 3~\textmu W for 100~ms, and the corresponding collected photon counts are used to assign the charge state following the model introduced in Ref.~\citenum{Shields2015}.

Figure~\ref{fig:sup}(a) shows an example of the photon distributions obtained during charge state readout. The green points show the experimentally observed photon probability distribution after a 515~nm pulse, which prepares NV$^-$ with $\sim 70 \%$ probability \cite{Waldherr2011} and red points show the photon counts after a 638~nm pulse, which prepares NV$^0$ with $\sim 90 \%$ probability \cite{Aslam2013}. The sum of the two measurements are given by the blue points, and the solid lines are fits to the model \cite{Shields2015}. From the model we can extract the electron photogeneration, hole photogeneration, NV$^0$ fluorescence, and NV$^-$ fluorescence rates for an individual NV center. The bimodal distribution of the green points illustrates the imperfect polarization into NV$^-$ after preparation under green illumination.

Figure~\ref{fig:sup}(b) shows the theoretical photon distributions of purely the NV$^-$ and NV$^0$ states, using the rates found from the blue curve in (a) and the model from \citeauthor{Shields2015}. The two charge states give distinct Poisson distributions. The threshold value that maximizes the fidelity of assigning the correct charge state to a single-shot measurement is represented by the vertical red line. For this measurement, the optimized threshold value of 8 counts gives a fidelity of $90\%$ in single-state charge state readout. For the measurements in the main text and supplemental material, our single-shot readout fidelity ranges from $80-90\%$. Figure~\ref{fig:sup}(c) shows an example of a single-shot measurement on the left, with the dark ring due to hole capture already weakly visible. In order to overcome shot noise, 20 to 30 single-shot measurements are taken consecutively, charge states are assigned for each respective spatial point for each measurement, and the single-shot charge state values are averaged together for each respective point (the averaged measurement is shown on the right).

\section{SiV charge photogeneration rates} \label{measuring_siv_ion}
\begin{figure}
\centering
\includegraphics[width=1.0\textwidth]{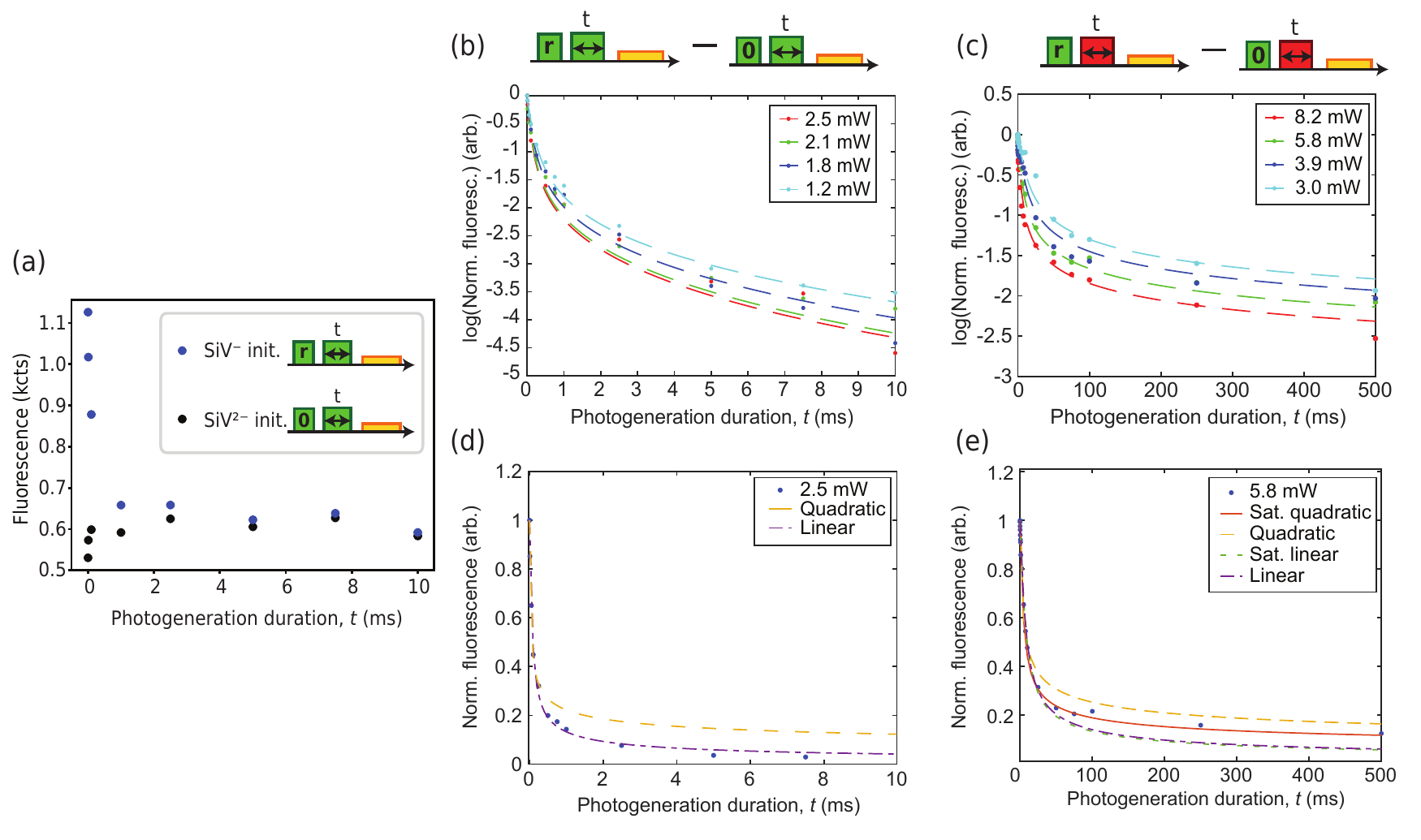}
	\caption{Measuring hole photogeneration rates $j_-$ of SiV centers. (a) Measurement to observe SiV hole photogeneration and conversion from SiV$^-$ to SiV$^{2-}$, with either SiV$^{-}$ initialization (blue dots) and SiV$^{2-}$ initialization (black dots).
	(b) Measurement of SiV$^-$ population decay from a 515 nm photogeneration pulse at various powers. Data is normalized by subtracting SiV$^{-}$ initialization data and SiV$^{2-}$ initialization data (see (a)), then dividing by the subtracted value at $t=0$ ms. The fit curves use a linear power dependence of $j_-$.
	(c) Measurement of SiV$^-$ population decay from a 638 nm photogeneration pulse at various powers. The fit curves use a saturated quadratic power dependence of $j_-$. 
	(d) $j_-$ power dependence comparison for 515 nm photogeneration pulse, using 2.5 mW data from (b). The model assuming $j_-$ has a pure linear power dependence best fits to the data. 
	(e) $j_-$ power dependence comparison for 638 nm photogeneration pulse, using 5.8 mW data from (c). The model assuming $j_-$ has a saturated quadratic power dependence best fits to the data. }
\label{fig:SiV_photogen_rates}
\end{figure}

In order to reduce the number of free parameters in the theoretical model described in the following section, we measured the charge  photogeneration rates of both NV centers and SiV centers under 515 nm and 638 nm illumination. In this section, we describe our method of determining the scaling with power of the photogeneration rates of the SiV centers under 515 nm (green) and 638 nm (red) illumination. 
Figure~\ref{fig:SiV_photogen_rates}(a) shows the measurements performed to determine the rate of hole photogeneration of SiV$^{-}$ to SiV$^{2-}$. The SiV centers at $r=0$ are either initialized in SiV$^-$ or SiV$^{2-}$. To initialize in SiV$^-$, a green laser pulse is positioned a few microns away from the SiV centers at $r=0$, which was demonstrated in the main text (see $\tau = 10$ ms data in Fig. 2(a)). Specifically, a 2~mW, 515 nm laser pulse is applied for 10~ms positioned 2.5~\textmu m away from the center, which is represented by the green pulse with an ``\textbf{r}". To initialize in SiV$^{2-}$, the green pulse is positioned directly on the SiV centers at $r=0$ for 10 ms, and is indicated by the green pulse with a ``\textbf{0}" on it. The photogeneration pulse (in this case, a 515 nm laser at 2.5 mW) is then applied at $r=0$ for a time $t$. Finally, the fluorescence of the SiV centers at $r=0$ are measured with a 0.4~mW, 40~ms 589~nm (yellow) readout pulse. This is repeated for different times of the photogeneration pulse, $t$.

This data is then normalized, and the normalized data for both 515 nm and 638 nm photogeneration pulses at various powers are shown in Figs.~\ref{fig:SiV_photogen_rates}(b) and (c) on a semilog scale. To normalize the data, the measurements of SiV$^{2-}$ initialization is subtracted form the measurement with SiV$^-$ initialization, as indicated above the plots, and then normalized to the value at $t=0$. Also shown in these plots are fits to the model described below.

Because of the high density of SiV centers in this sample, the SiV centers within the laser beam's profile experience different intensities of light, thus optically converting to SiV$^{2-}$ at different rates, and so the measurement captures all these different rates, as is seen in the the super-exponential decays in Fig.~\ref{fig:SiV_photogen_rates}(b) and (c). To accurately model this process, we consider the laser beam's influence over an evenly distributed density of SiV centers. The laser profile and the collection efficiency profile by our confocal microscope are both Gaussian beam profiles. We integrate over these profiles in three dimensions to extract the fluorescence from SiV$^-$ as a function of photogeneration pulse duration and the photogeneration rates. 

Through the measurements presented in Fig. 2(a) in the main text and Fig.~\ref{fig:red_ionize}, we observe that direct illumination with either 515 nm and 638 nm light converts SiV$^-$ to SiV$^{2-}$ through hole photogeneration, and the only way for the defect to return to SiV$^-$ is through capturing a hole photogenerated form other surrounding defects.
Because of this, we assume $j_{2-}$, the electron photogeneration rate of SiV$^{2-}$ to SiV$^-$, is insignificant compared to $j_{-}$, the hole photogeneration rate of SiV$^{-}$ to SiV$^{2-}$, and thus $j_{2-}$ is taken to be 0 (Table~\ref{table:params}). In addition, we do not know if the photogeneration processes are linear or quadratic, so we therefore fit to models describing both.

Figure~\ref{fig:SiV_photogen_rates}(d) shows fits with linear or quadratic power scaling of $j_-$ to the 2.5 mW, 515 nm photogeneration pulse data (red circles and fits in (b)). A linear power dependence shows the best agreement with the data, indicating that 515 nm illumination optically converts SiV$^-$ to SiV$^{2-}$ through a one-photon process, consistent with prior results \cite{Dhomkar2018}. The power dependent photogeneration rate of SiV$^-$ under 515 nm illumination is found to be: 
\begin{equation}
{j_-}^{515} = v_g \times I,
\end{equation}
where $v_g = 1.02 \times 10^3$ s$^{-1}$/(mW/(\textmu m)$^2)$ (see Table~\ref{table:params}) and $I$ is in units of mW/\textmu m)$^2$. 

Figure~\ref{fig:SiV_photogen_rates}(e) shows fits with various power scaling of $j_-$ to the 5.8 mW, 638 nm photogeneration pulse data (green circles and fits in (c)). A quadratic power dependence accounting for saturation shows the best agreement with the measurement, suggesting that under 638 nm illumination the optical conversion from SiV$^-$ to SiV$^{2-}$ is a two-photon process. We note that prior work \cite{Nicolas2019} reports a two-photon process with 737 nm wavelength. The obtained hole photogeneration rate of SiV$^-$ under 638 nm illumination is: 
\begin{equation}
{j_-}^{638} = v_r\times \frac{I^2}{1 +I/I_{sat} },
\end{equation}
where $v_r =  4.3 $ s$^{-1}$/(mW/(\textmu m)$^2)$ and $I_{sat} = 4.8$ mW/(\textmu m)$^2$. 

\section{Modeling SPaCE measurements}

\begin{figure}
\centering
\includegraphics[width=1.0\textwidth]{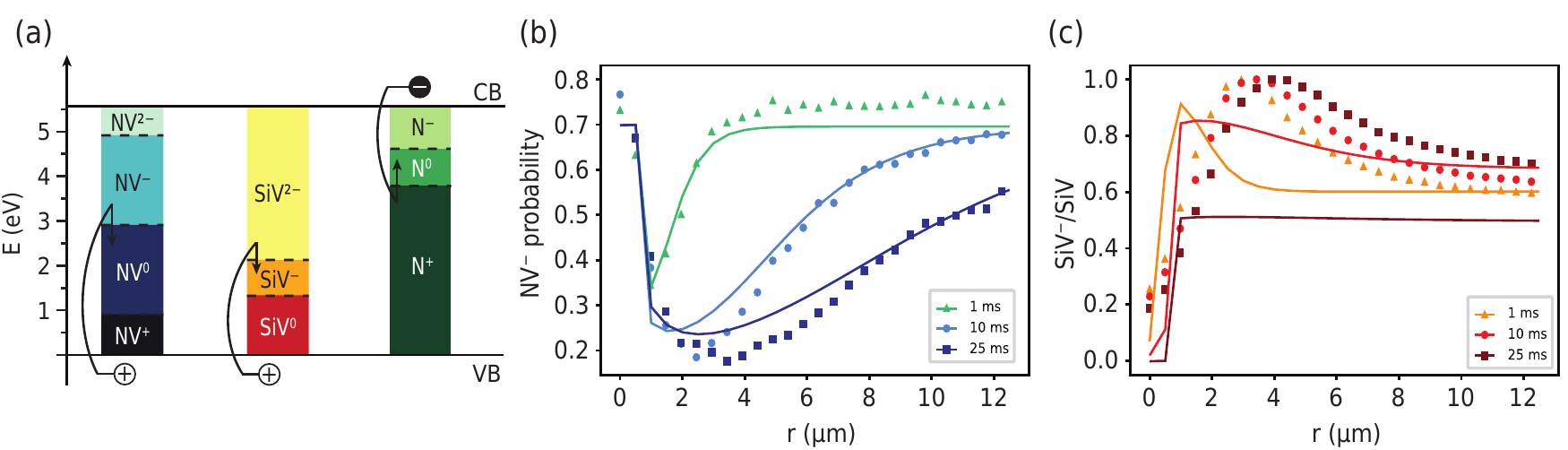}
	\caption{Modeling SPaCE measurements. (a) Energy band diagram of diamond showing the charge transition levels of the NV center \cite{Deak2014}, SiV center \cite{Gali2013}, and N defect \cite{Deak2014}. Cartoon of holes from the valence band (VB) being captured by NV$^-$ and SiV$^{2-}$ are shown, as well as an electron from the conduction band (CB) being captured by N$^+$. (b) Radial average data from NV SPaCE measurements for CPG pulse times 1 ms (teal triangles), 10 ms (light blue circles), and 25 ms (dark blue squares), with modeled curves using parameters from fitting to this data for each respective time. (c) Radial average data from SiV SPaCE measurements, after normalizing counts to population, for CPG pulse times 1 ms (yellow triangles), 10 ms (light red circles), and 25 ms (dark red squares), with modeled curves using parameters from fitting to NV data for each respective time. }
\label{fig:charge_model}
\end{figure}

In this section we give details of the theoretical model used to attempt to simulate the charge dynamics observed in our experiments. We note, however, we were not able to converge on a single set of parameters to quantitatively simulate measurements. We adapt the theoretical description used in Refs.~\citenum{Dhomkar2018}, \citenum{Jayakumar2016}  which models the cylindrically symmetric diffusion and capture of charge carriers (electrons and holes) that are photogenerated by defects under optical illumination. Figure~\ref{fig:charge_model}(a) illustrates some of the processes in this model, like hole and electron capture, 
of the relevant defects in our sample. Shown are the charge transition levels of the defects, which depict the Fermi level position for which the formation energies of the defect in the two charge states are equal \cite{Gali2013}. These levels are related to the optical absorption energies required to photoexcite carriers into or out of the defects. 
The colored regions indicate which charge state of each defect is most stable thermodynamically for given value of the Fermi level energy. The Fermi level in our diamond sample lies close to the middle of the band gap, so the stable charge states are NV$^-$, SiV$^{2-}$, and N$^+$, which also follow from application of the octet rule for a neutral crystal. The corresponding excited states lie in the adjacent near-mid gap regions, indicating the relevant charge transitions to be NV$^-$/NV$^0$, N$^+$/N$^0$, and SiV$^{2-}$/SiV$^-$. 
Based on these observations, we only consider the NV$^{-}$/NV$^0$, N$^+$/N$^0$, and SiV$^-$/SiV$^{2-}$ charge states in our theoretical model.

The theoretical model from Refs.~\citenum{Dhomkar2018}, \citenum{Jayakumar2016}  is described by a set of coupled differential equations, which we adapt and show in Eq.~\ref{eq:pdes}:
\begin{eqnarray}\label{eq:pdes}
&&\frac{\partial Q_-}{\partial t}=-(k_-+\kappa_p p)Q_- + (k_0+\kappa_n n)Q_0, \label{eq_Qn}\nonumber \\
&&\frac{\partial Q_0}{\partial t}=(k_-+\kappa_p p)Q_- - (k_0+\kappa_n n)Q_0, \nonumber \\
&&\frac{\partial P_0}{\partial t}=\gamma_n n P-(k_N+\gamma_n n+\gamma_p p)P_0, \nonumber \\
&&\frac{\partial S_-}{\partial t}=\left(j_{2-}+\chi_p p \right)S-\left(j_{-}+j_{2-}+\chi_n n+\chi_p p\right) S_-, \nonumber\\
&&\frac{\partial n}{\partial t}= D_n \nabla^2 n +k_- Q_--\kappa_n n Q_0+k_N P_0-\gamma_n n \left(P-P_0\right)+j_{2-} (S-S_{-}) -\chi_n n S_{-},\nonumber\\
&&\frac{\partial p}{\partial t}= D_p \nabla^2 p -\kappa_p p Q_- + k_0 Q_0 -\gamma_p p P_0 +j_- S_{-} -\chi_p p (S-S_{-}), \nonumber\\
\end{eqnarray}
where $Q_-$, $Q_0$, $P_0$, and $S_-$ are the density of NV$^-$, NV$^0$, N$^0$, SiV$^-$ states, respectively, and $P$ and $S$ are the total concentration of N and SiV defects, respectively. The last two equations describe the change in $n$ and $p$, the local densities of the free electrons and holes, respectively. The other variables are defined in Table~\ref{table:params}. The system of equations (Eq.~\ref{eq:pdes}) is obtained by taking into account that the total number of NV, SiV, and N defects are conserved, given by 
\begin{equation}
    Q=Q_0+Q_-, \qquad P=P_0+P_+, \qquad S=S_-+S_{2-},
\end{equation}
where $P_+$, and $S_{2-}$ are the densities of N$^+$, and SiV$^{2-}$, respectively. The first two equations of Eq.~\ref{eq:pdes} serve numerically to enforce charge conservation of a single NV center at the center of the simulation domain. In the above system of equations, as an approximation, the electric field effect due to the distribution of charges is excluded. 

\begin{table*}
	\caption{Parameters given in the system of partial differential equations, Eq.~\ref{eq:pdes}. Values marked with [$\star$] represents the parameters we have measured or observed, while values with [$\dagger$] are the parameters we have fit to the SPaCE measurements given in Fig.~\ref{fig:charge_model}(b) (see Table~\ref{table:fitparams}). The variable $r$ is the radial distance of the laser illumination from the center $r=0$.}\label{table:params}
	\vspace{0.25cm}
	\begin{tabular}{|l|*{5}{c|}}\hline
		\textbf{Parameter (units)} & \textbf{Description} & \textbf{Value}  \\
			\hline
			\hline 
			$w_0$~(\textmu m) & Laser beam 1/e$^2$ radius  &   [$\dagger$] \Tstrut\\
			& 515 nm&  \Bstrut\\
			\hline
			$\mathcal{P}$ (mW) & Laser beam power, 515 nm & 2 [$\star$]  \TBstrut \\
			\hline
			$P$ (defect/\textmu  m$^3$) & N density & [$\dagger$] \TBstrut \\
			\hline
			$S$ (defect/\textmu  m$^3$) & SiV density & [$\dagger$] \TBstrut\\
			\hline 
			$D_n$~(\textmu m$^2$/s) & Free electron diffusion constant & $5.5 \times 10^9 $ [\citenum{Han1995}] \TBstrut\\
			\hline 
			$D_p$~(\textmu m$^2$/s) & Free hole diffusion constant & $4.3 \times 10^9$ [\citenum{Han1995}]  \TBstrut\\
			\hline

			$k_0 ~(\times 10^6$ s$^{-1}$) & NV$^0$ hole photogeneration rate  &  $k_0=5.173\times\frac{4 \mathcal{P}^2}{\pi w^2_0} e^{(-4(r-r_0)^2/w^2_0)}$ [$\star$] \Tstrut\\ 
			&under 515 nm illumination& \Bstrut\\
			\hline
			$k_-~(\times 10^6$ s$^{-1}$) & NV$^-$ electron photogeneration rate  & $k_-=2.217\times\frac{4 \mathcal{P}^2}{\pi w^2_0} e^{(-4(r-r_0)^2/w^2_0)}$  [$\star$] \Tstrut\\
			&under 515 nm illumination& \Bstrut\\
			\hline 
			$k_N$ (s$^{-1}$)&  N$^0$ electron photogeneration rate &   $k_N=\tilde k_N \times e^{(-2(r-r_0)^2/w^2_0)}$  [$\dagger$] \Tstrut\\
			&under 515 nm illumination& $(\tilde k_N$ is the fit parameter)\Bstrut\\
			\hline 
			$j_{2-}$ (s$^{-1}$ ) &  SiV$^{2-}$ electron photogeneration rate & 0 [$\star$] \Tstrut\\
			&under 515 nm illumination&\Bstrut\\
			\hline 
			$j_{-} ~(\times 10^3$ s$^{-1}$) &  SiV$^-$ hole photogeneration rate &$j_{-}=1.02\times\frac{2 \mathcal{P}^2}{\pi w^2_0} e^{(-2(r-r_0)^2/w^2_0)}$  [$\star$]\Tstrut \\
			&under 515 nm illumination& \Bstrut\\
			\hline

			$\kappa_n$ (s$^{-1}$~\textmu  m$^3)$ & NV$^0$ electron capture rate & 0 [\citenum{Dhomkar2018}, $\star$] \TBstrut\\
			\hline
			$\kappa_p$ (s$^{-1}$~\textmu  m$^3)$ & NV$^-$ hole capture rate  & [$\dagger$] \TBstrut\\
			\hline
			$\gamma_n$ (s$^{-1}$~\textmu  m$^3)$ & N$^+$ electron capture rate  &   [$\dagger$]  \TBstrut\\
			\hline 
			$\gamma_p$ (s$^{-1}$~\textmu  m$^3)$ & N$^0$ hole capture rate &  [$\dagger$] \TBstrut\\
			\hline
			$\chi_n$ (s$^{-1}$~\textmu  m$^3)$ & SiV$^-$ electron capture rate & [$\dagger$] \TBstrut\\
			\hline 
			$\chi_p$ (s$^{-1}$~\textmu  m$^3)$ & SiV$^{2-}$ hole capture rate & [$\dagger$]  \TBstrut\\
			\hline
		\end{tabular}
\end{table*}

To satisfy the charge neutrality for the initial densities of the defects, we impose $P_{+}-2S_{2-}-S_{-}-Q_{-}=0$. Moreover, we take the initial densities of free electrons and holes to be zero, which is reasonable for the very wide bandgap of diamond. The boundary conditions are obtained by imposing the charge conservation in the sample, i.e.,
\begin{equation}
    \int{\rho dV}=0, \qquad \rho=e\left({P_{+}+p-n-2S_{2-}-S_{-}-Q_{-}}\right).
\end{equation}
Using the system of equations (Eq.~\ref{eq:pdes}) and Gauss' theorem to convert the volume integral to surface integral we obtain
\begin{equation}
  \nabla n\cdot \vec s|_s=0, \quad \nabla p \cdot \vec s|_s=0,
\end{equation}
where $s$ is the surface enclosing the volume of the sample and $\vec s$ is the outward normal unit vector to the surface. 

We model the SPaCE measurements \cite{Jayakumar2016, Dhomkar2018} with a discrete NV density such that there is only one NV at the center, i.e. at $r=0$, while still including continuous densities of SiV and N defects (taken as free parameters in the model). We first simulate the initialization pulse at $r=0$, followed by the CPG pulse at position $r$. The effect of both laser pulses is simulated with a Gaussian profile, with a 1/e$^2$ width $w_0$ (Table \ref{table:params}), centered at the point of illumination $r$, as shown in Table~\ref{table:params}, the photogeneration rates associated with two-photon (one-photon) processes are taken to scale quadratically (linearly) with laser power. Following the original model\cite{Jayakumar2016, Dhomkar2018}, cylindrical symmetry is assumed for the system, which significantly simplifies numerical simulations. The resulting charge carrier photogeneration from defects and diffusion of carriers due to the CPG pulse are modeled based off of Eq.~\ref{eq:pdes} in the radial dimension $r = 100$ \textmu m
. Matlab pdepe solver was used, which uses method of lines together with a finite element discretization in space \cite{Skeel1990}. Finally, the NV$^-$ probability of the single NV at the center, as well as the SiV$^-$ density at the center, are extracted from the simulations as a function of distance between the initialized NV center and CPG pulse.

\begin{table*}
	\caption {Values of the parameters from fits to the radially averaged NV SPaCE measurements for different CPG pulse times, see Fig.~\ref{fig:charge_model}(b).}\label{table:fitparams}
	\vspace{0.25cm}
	\begin{tabular}{|l|c|c|c|}\hline
		    &\multicolumn{3}{c|}{\textbf{CPG Pulse Time}}  \\
			\hline
		    \textbf{Parameter (units)} & \textbf{1 ms} & \textbf{10 ms} & \textbf{25 ms} \\
			\hline
			\hline
			$w_0~($\textmu  m) &0.4&0.33&0.4 \TBstrut\\
			\hline
			$P$ (defect/\textmu  m$^3$) & 2300 & 1896 & 2500 \TBstrut \\
			\hline
			$S$ (defect/\textmu  m$^3$) & 4933.6 & 1444 & 1150   \TBstrut\\
			\hline
			SiV$^-$ initial density (defect/\textmu  m$^3$) & $0.6 S$ & $0.7 S$& $0.5 S$  \TBstrut  \\
			\hline
			$\tilde k_N$ (s$^{-1}$) & $30\times 10^3$ & $34\times 10^3$ &  $69.6\times 10^3$ \TBstrut\\
			\hline
			$\kappa_p$ (s$^{-1}$~\textmu  m$^3)$ & $8.1\times 10^6$ & $8.4\times 10^6$ & $7.2\times 10^6$   \TBstrut\\
			\hline
			$\gamma_n$ (s$^{-1}$~\textmu  m$^3)$ & 100.77 & 280.06 &  990 \TBstrut \\
			\hline
			$\gamma_p$ (s$^{-1}$~\textmu  m$^3)$ & $2.99 \times 10^5$  & $2.67\times10^5$ &  $4\times 10^5$  \TBstrut\\
			\hline
			$\chi_n$ (s$^{-1}$~\textmu  m$^3)$ & 104.5 & 98.6 &  306.1  \TBstrut\\
			\hline
			$\chi_p$ (s$^{-1}$~\textmu  m$^3)$ & $8.68\times 10^6$ & $2.93\times 10^6$ & $10^5$  \TBstrut \\
			\hline
		\end{tabular}
\end{table*}

Figure~\ref{fig:charge_model}(b) shows the results of fitting this model to the NV SPaCE measurements for three CPG pulse times, where the free parameters (see Table~\ref{table:params}) of the fits are found for each measurement separately. Those fit parameters are then used to simulate the response from an ensemble of SiV centers at $r =0$ within the confocal volume, where the SiV$^-$ fractional population at $r=0$ is plotted in Fig.~\ref{fig:charge_model}(c) (the SiV experimental data was normalized to the peak value for each respective measurement). We were not able to find agreement in the fit parameters across all three CPG pulse times, especially the SiV$^{2-}$ hole capture rate $\chi_p$ (see Table \ref{table:fitparams}). We note that the NV$^-$ hole capture rate, $\kappa_p$ was consistently found to be about three orders of magnitude larger than previously reported \cite{Jayakumar2016, Dhomkar2018}, suggesting that the NV center is more susceptible to hole capture than expected. The simulated NV curves qualitatively describe the data, however quantitative agreement was not achieved. In addition, the modeled SiV curves are not in good agreement with the experimental data. We hypothesize that expanding this model to consider out-of-plane effects of laser illumination and diffusion, and thus solve the system of equations (Eq.~\ref{eq:pdes}) in three dimensions, may result in better agreement between theory and experiment.

\bibliography{MainReferences}